\documentclass[aps,pre,unsortedaddress,twocolumn,showpacs,showkeys]{revtex4}

\usepackage[latin1]{inputenc}

\usepackage{graphicx}
\usepackage{color}
\usepackage{hyperref}

\newcommand{\incfig}[2] {\includegraphics[width=#1]{#2}}

\newcommand{\outline}[1]{}

\begin{document}

\title{Compounding approach for univariate time series with non-stationary variances}

\author{Rudi Sch\"{a}fer}
\email[]{rudi.schaefer@uni-duisburg-essen.de}
\affiliation{Fakult\"{a}t f\"{u}r Physik, Universit\"{a}t Duisburg-Essen, Germany}

\author{Sonja Barkhofen}
\email[]{sonja.barkhofen@uni-paderborn.de}
\affiliation{Applied Physics, University of Paderborn, Warburger Strasse 100, 33098 Paderborn, Germany}

\author{Thomas Guhr}
\email[]{thomas.guhr@uni-duisburg-essen.de}
\affiliation{Fakult\"{a}t f\"{u}r Physik, Universit\"{a}t Duisburg-Essen, Germany}

\author{Hans-J\"{u}rgen St\"{o}ckmann}
\email[]{stoeckmann@physik.uni-marburg.de}
\affiliation{Fachbereich Physik, Phillips-Universit\"{a}t Marburg, Germany}

\author{Ulrich Kuhl}
\email[]{ulrich.kuhl@unice.fr}
\affiliation{Laboratoire de Physique de la Mati\`{e}re Condens\'{e}e, CNRS UMR 7336, Universit\'{e} de Nice Sophia-Antipolis, F-06108 Nice, France}

\date{\today}

\begin{abstract}
A defining feature of non-stationary systems is the time dependence of their statistical parameters. 
Measured time series may exhibit Gaussian statistics on short time horizons, due to the central limit theorem. The sample statistics for long time horizons, however, averages over the time-dependent parameters.
To model the long-term statistical behavior, we compound the local distribution with the distribution of its parameters.
Here we consider two concrete, but diverse examples of such non-stationary systems, the turbulent air flow of a fan and a time series of foreign exchange rates. Our main focus is to empirically determine the appropriate parameter distribution for the compounding approach. To this end we have to estimate the parameter distribution for univariate time series in a highly non-stationary situation.
\end{abstract}

\maketitle

\section{Introduction}   \label{sec1}

Due to the central limit theorem a great deal of phenomena can be described by Gaussian statistics. This also guides our perception of the risks of large deviations from an expectation value.
Consequently, the occurence of any aggravated probability of extreme events is always cause for concern and subject of intense research interest. 
In a large variety of systems where heavy-tailed distributions are observed, Gaussian statistics holds only locally --- the parameters of the distribution are changing, either in time or in space. Thus, to describe the sample statistics for the whole system, one has to average the parametric distribution over the distribution of the (shape) parameter.
This construction is known as \textit{compounding} or \textit{mixture}
\cite{dubey1970compound,Barndorff-Nielsen1982,Doulgeris2010} in the
mathematics and as \textit{superstatistics} \cite{Beck03} in the
physics literature.

An important example for parameter distribution functions is the K-distribution mentioned 1978 for the first time by Jakeman and Pusey \cite{jak78}. It was introduced in the context of intensity distributions, and their significance for scattering processes of a wide range of length scales was stressed.
Moreover the distribution is known to be an equilibrium solution for the population in a simple birth-death-immigration process which was already applied in the description of eddy evolution in a turbulent medium. The underlying picture of turbulence assumes that large eddies are spontaneously created and then ''give birth`` to generations of children eddies, which terminates when the smallest eddies die out due to viscous dissipation.
In \cite{jak76} Jakeman and Pusey use the K-distribution for fitting data of microwave sea echo, which turned out to be highly non-Rayleigh.
The K-distribution is also found as a special case of a full statistical-mechanical formulation for non-Gaussian compound Markov process, developed in \cite{jak88}.
Field and Tough find K-distributed noise for the diffusion process in electromagnetic scattering \cite{fie03, fie03b}.
Experimentally the K-distribution appeared in the contexts of irradiance fluctuations of a multipass laser beam propagating through atmospheric turbulence \cite{maj82},  synthetic aperture radar data \cite{lee94}, ultrasonic scattering from tissues \cite{sha95,wen91} and mesoscopic systems \cite{arn09}.
Also in our study we will encounter the K-distribution for one of the systems under consideration.

Compounded distributions can be applied to very different empirical situations:
They can describe aggregated statistics for many time series, where each time series obeys stationary Gaussian statistics, the parameters of which vary only between time series.
In this case it is straightforward to estimate the parameter distribution.
The situation is more difficult when we consider the statistics of single long time series with time-varying parameters. In this non-stationary case, one often makes an ad hoc assumption about the analytical form of the parameter distribution, and only the compounded distribution is compared to empirical findings.

In this paper we address the problem of determining the parameter distribution empirically for univariate non-stationary time series. Specifically we consider the case of Gaussian statistics with time-varying variance. In this endeavor we encounter several problems: If the variance for each time point is purely random, as the compouding ansatz would suggest, we have no way of determining the variance distribution from empirical data.
A prerequisite for an empirical approach to the parameter distribution is a time series which is quasi-stationary on short time intervals. In other words, the variance should vary only slowly compared to the time scale of fluctuations in the signal.
The estimation noise for the local variances competes with the variance distribution itself.
Therefore the time interval on which quasi-stationarity holds, should not be too short.
Furthermore, we have to heed possible autocorrelations in the time series themselves, since they might lead to an estimation bias for the local variances.
Our aim is to test the validity of the compounding approach on two different data sets.

The paper is organized as follows: In section \ref{sec2} we give a short summary of the compounding approach and present two recent applications where the K-distribution comes into play. In section \ref{sec3} we introduce the two systems we are going to analyse, a table top experiment on air turbulence and the empirical time series of exchange rates between US dollar and Euro. In section \ref{sec4} we address the problem of estimating non-stationary variances in univariate time series. Our empirical results are presented in section \ref{sec5}.

\section{Theoretical considerations} \label{sec2}

We consider a distribution $p(x|\alpha)$ of $d$ random variables,
ordered in the vector $x$. It is also a function of a parameter
$\alpha$ that determines the shape or other features of the
distribution, e.g. the variance of a Gaussian. If, in a given data set, the parameter $\alpha$ varies
in an interval $A$, one can try to construct the distribution of $x$
as the linear superposition
\begin{equation}
\langle p\rangle(x) = \int\limits_A f(\alpha) p(x|\alpha) d\alpha
\label{comp1}
\end{equation}
of all distributions $p(x|\alpha)$ with $\alpha\in A$. Here,
$f(\alpha)$ is the weight function determining the contribution of
each value of $\alpha$ in the superposition. Since $\alpha$ itself
typically is a random variable, we assume that the function $f(\alpha)$
is a proper distribution. In particular, it is positive semidefinite. 
As each $p(x|\alpha)$ and the resulting $\langle p\rangle(x)$ have to
be normalized with respect to the random vector $x$,
Eq.~(\ref{comp1}) implies the normalization
\begin{equation}
\int\limits_A f(\alpha) d\alpha = 1 .
\label{comp2}
\end{equation}

\begin{figure}
\begin{center}
 \incfig{0.46\textwidth}{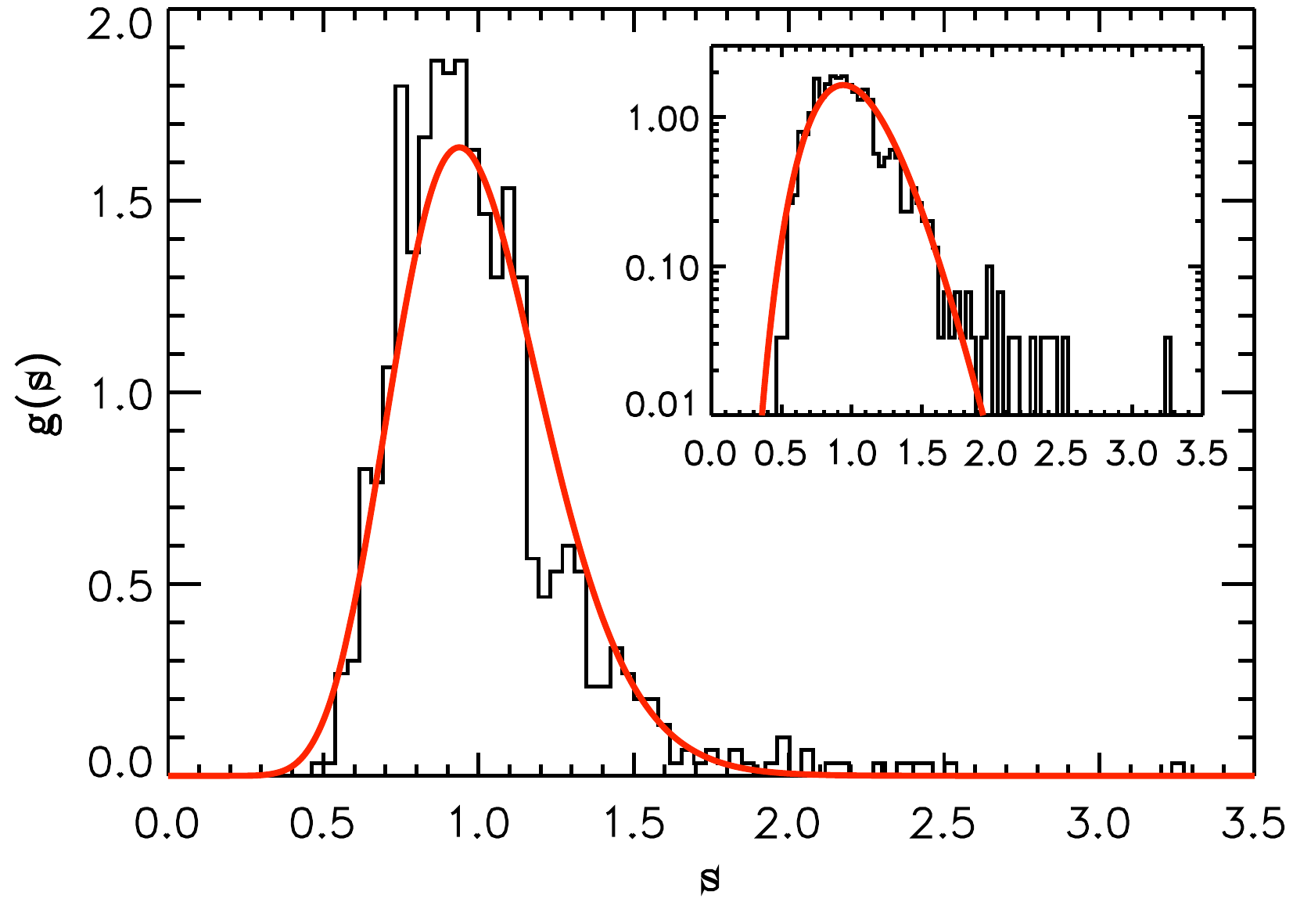}\\[2ex]
 \incfig{0.46\textwidth}{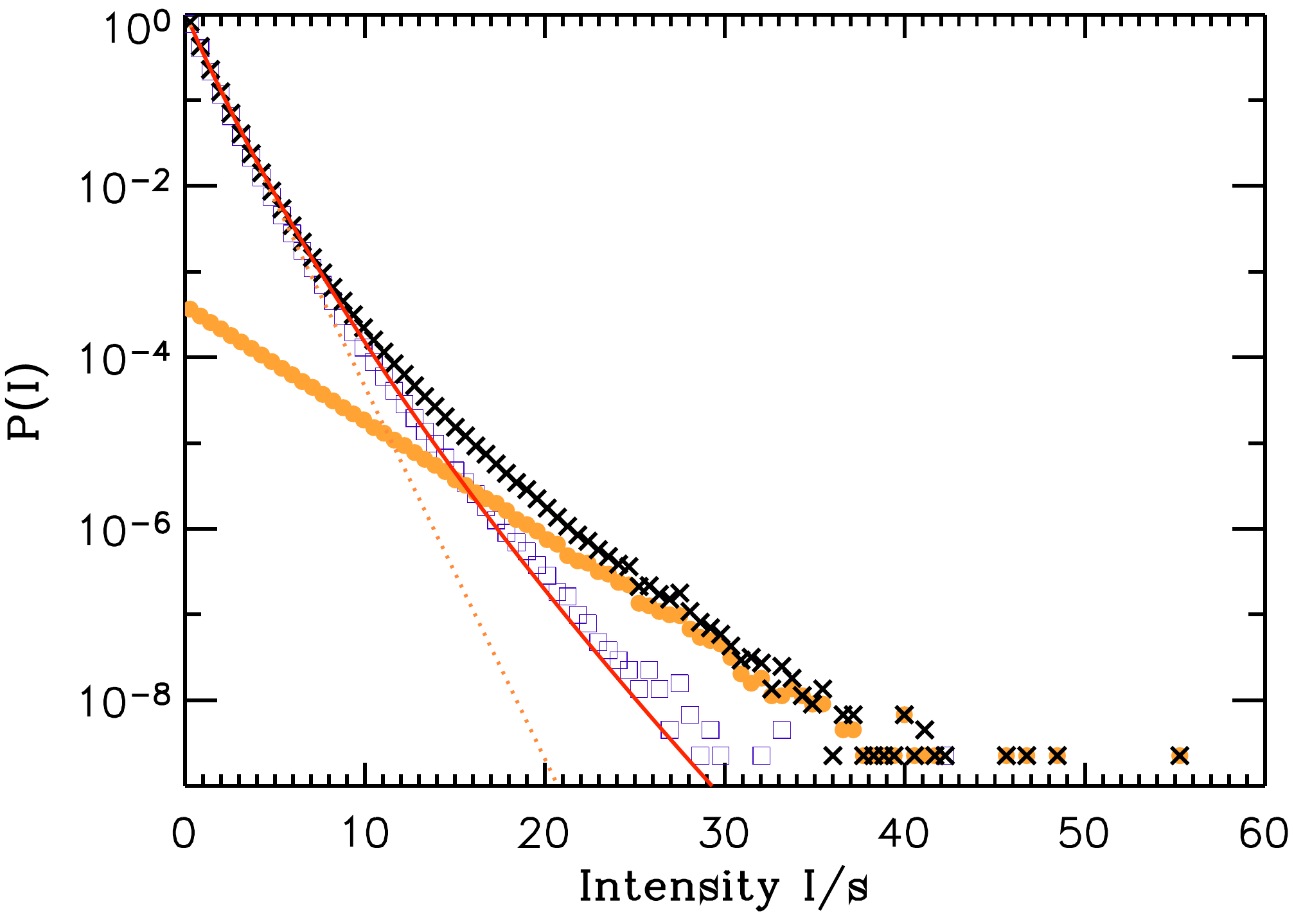}
\caption{(Top) Distribution of the time-averaged intensities $I_\textrm{loc}$ found for the 780 pixels of our measurement. The inset shows the same data using a semi-logarithmic scale. The solid curve is a $\chi^2$ distribution with $\nu=32$ degrees of freedom.
(Bottom) Intensity distribution for the time-dependent wave patterns generated by Eq.~(\ref{eq:pulse}). The solid (red) line is given by Eq.~(\ref{eq:kbess}).}
\label{fig:freak}
\end{center}
\end{figure}

The physics reasons for the variation of the parameter $\alpha$ can be
very different. In non--equilibrium thermodynamics, $\alpha$ might be
the locally fluctuating temperature.
Although our systems are not of a 
thermodynamic kind, we also have in mind non--stationarities.  In
recent experiments \cite{hoeh10,bar13c}  we studied the propagation of microwaves
through an arrangement of disordered scatterers in a cavity. The distribution of the electric fields was measured at fixed frequencies as a function of position. Then time-dependent wave fields were generated by superposition of $N=150$ patterns,
\begin{equation}\label{eq:pulse}
  \psi(x,y,t) = \sum_{i=1}^N\psi_i(x,y)e^{\imath (2 \pi f_i t-\varphi_i)}.
  \label{eq::TransientDef}
\end{equation}
Here $\psi_i(x,y)$ is the wave pattern at frequency $f_i$,  and $\varphi_i$ is a random phase. For fixed positions, always a Rayleigh distribution  was found in the time sequence $\psi(x,y,t)$ for the distribution of intensities $I$,

\begin{equation}\label{eq:raleigh}
    p(I|I_\textrm{loc})=\frac{1}{I_\textrm{loc}}\exp(-I/I_\textrm{loc})\,.
\end{equation}
 This is nothing but a manifestation of the central limit
theorem. The variance, \textit{i.e.}, the averaged
$I_\textrm{loc}$ depends on the position.  The large amount of data
made it possible to extract the distribution of the parameter
$I_\textrm{loc}$.  In good approximation, it turned out to be a $\chi_\nu^2$
distribution, see Fig. \ref{fig:freak}. The number $\nu$ of degrees of freedom was related to the
number of independent field components and took a value of $\nu=30$.
The authors of Refs.~\cite{hoeh10,bar13c} then used the compounding ansatz
Eq.~(\ref{comp1}) in the form
\begin{equation}
\langle p\rangle(I) = \int\limits_0^\infty \chi_\nu^2(I_\textrm{loc}) p(I|I_\textrm{loc}) dI_\textrm{loc} \ ,
\label{comp3}
\end{equation}
The integral can be done and yields
\begin{equation}
\langle p\rangle(I)=\frac{\nu}{\Gamma\left(\frac{\nu}{2}\right)}
       \left(\frac{\nu I}{2}\right)^{\frac{\nu}{4}-\frac{1}{2}} \mathcal{K}_{\frac{\nu}{2}-1}\left(2\sqrt{\frac{\nu I}{2}}\right) \ ,
\label{eq:kbess}
\end{equation}
where $\mathcal{K}_\mu$ is the modified Bessel function of degree
$\mu$. This is the K-distribution introduced in the introduction.  Omitting local regions with extremely high amplitudes, so--called ``hot spots'', the intensity distributions could be perfectly well interpreted in terms of K-distributions, see Fig. \ref{fig:freak}.

We briefly sketch another example which stems from finance. Details
can be found in Ref.~\cite{Schmitt2013a}. The conceptual difference to
the previous example is that the compounding formula~(\ref{comp1})
appears as an intermediate result, not as an ansatz. We consider a
selection of $K$ stocks with prices $S_k(t), \ k=1,\ldots,K$ belonging
to the same market. One is interested in the distribution of the
relative price changes over a fixed time interval $\Delta t$,
also referred to as returns
\begin{equation}
 r_k (t) = \frac{S_k(t + \Delta t)-S_k(t)}{S_k(t)} \ .
\label{returns}
\end{equation}
As the companies operate on the same market, there are correlations
between the stocks which have to be measured over a time window $T$
much larger than $\Delta t$. The corresponding covariances form the
$K\times K$ matrix $\Sigma_t$. The business relations
between the companies as well as the market expectations of the stock
market traders change in time. Thus, the market is non--stationary and
the correlation coefficients fluctuate in time. Only for time windows
$T$ of a month or less, $\Sigma_t$ is approximately constant. The
multivariate distribution of the returns $r_k=r_k(t)$ ordered in the
$K$ component vector $r$ at a given time $t$ is well described by the
Gaussian
\begin{equation}
p(r|\Sigma_t)  = \frac{1}{\sqrt{\det(2\pi\Sigma_t)}}
                       \exp\left( -\frac{1}{2} r^\dagger \Sigma_t^{-1}r\right) \ .
\label{multivar}
\end{equation}
The non--stationarity over longer time windows $T$ can be modeled by observing
that the ensemble of the fluctuating covariance matrices $\Sigma_t$
can be approximated by an ensemble of random matrices.
In Ref.~\cite{Schmitt2013a} a Wishart distribution was assumed for this ensemble.
The ensemble average of the distribution~(\ref{multivar}) over the Wishart ensemble yields
\begin{equation}
\langle p \rangle (r|\Sigma,N) = \int\limits_0^\infty \chi_N^2(z) p\left(r\Big|\frac{z}{N}\Sigma\right) dz \ ,
\label{compound}
\end{equation}
where $\Sigma$ is the sample-averaged covariance matrix over the
entire time window $T$. As $\Sigma$ is fixed, this result has the form
of the compounding ansatz~(\ref{comp1}). Furthermore, it closely
matches the result~(\ref{comp3}) found in the context of microwave
scattering. The number $N$ of degrees of freedom in the $\chi_N^2$
distribution determines the variance in the distribution of the random
covariance matrices. The role of the locally averaged intensity
$I_\textrm{loc}$ is now played by an effective parameter $z$ which
fully accounts for the ensemble average. Again, a K-distribution follows, 
\begin{eqnarray}
\langle p \rangle (r|\Sigma,N) &=&
                \frac{1}{2^{N/2+1}\Gamma(N/2)\sqrt{\det(2\pi\Sigma/N)}} \nonumber \\
 &&         \times \ \frac{\mathcal{K}_{(K-N)/2}\left(\sqrt{Nr^\dagger\Sigma^{-1}r}\right)}
                                      {\sqrt{Nr^\dagger\Sigma^{-1}r}^{(K-N)/2}} \ ,
\label{ergebnis}
\end{eqnarray}
in which the bilinear form $r^\dagger\Sigma^{-1}r$ takes the place of the intensity
$I$ in Eq.~(\ref{eq:kbess}).
In both examples, averages over fluctuating quantities produce heavy-tailed distributions which describe the vast majority of the large events.

\section{Data acquisition} \label{sec3}
\subsection{Turbulent air flow}
\begin{figure}
\begin{center}
 \incfig{0.46\textwidth}{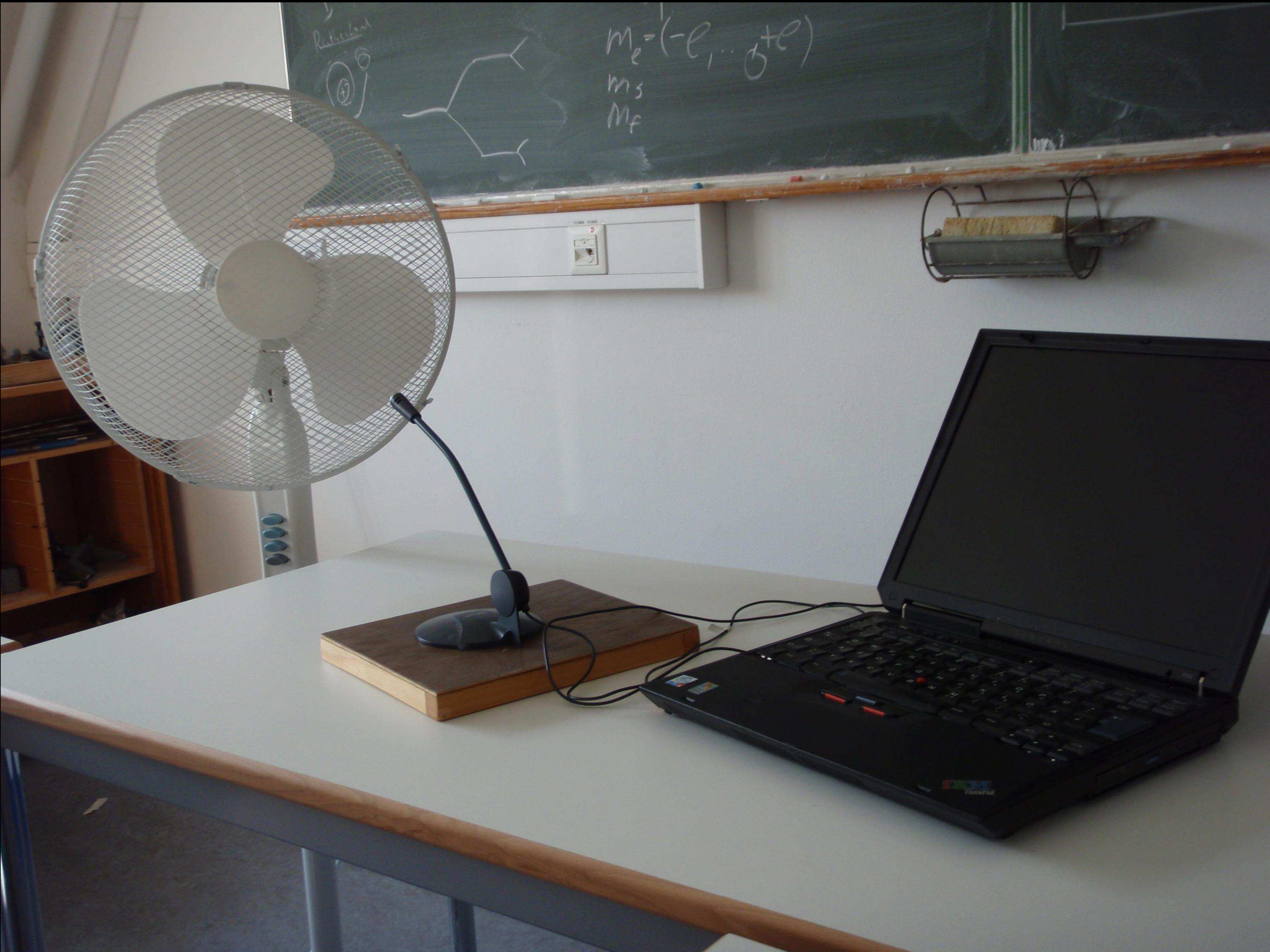}
\caption{Setup for the generation and measurement of the turbulent air flow. 
The turbulent air flow has been generated by the fan on the left hand side and recorded by the microphone in front of it. In the preliminary stage of the experiments the sound card of the laptop has been used for data acquisition.}
\label{fig:sound}
\end{center}
\end{figure}

The first data set is obtained by measuring the noise generated by a turbulent air flow.
For the turbulence generation we used a standard fan with a rotor frequency of 18.44\,Hz.
We restricted ourselves to standard audio technique handling frequencies up to 20\,kHz and standard sampling rates of 48\,kHz offering reliable quality at an attractive price.
The microphone for the sound recording is a E 614 by Sennheiser with a frequency response of 40\,Hz - 20\,kHz, a good directional characteristic and a small diameter of 20\,mm.
It guarantees a broadband frequency resolution and a point-like measuring position.
An external sound card with matching properties was necessary to use the full capacity of the quality of microphone and to minimize the influence of the intrinsic noise of the PC. 
Fig. \ref{fig:sound} shows a photograph or the used setup.

A microphone has been placed in front of a fan  running continuously and generating a highly turbulent air flow. The microphone records the time signal of the sound waves excited by the turbulence. 
The details about the analysed time series will be discussed in section \ref{sec4}.

\subsection{Foreign exchange rates}

The foreign exchange markets have the peculiar feature of all-day continuous trading. This is in contrast to stock markets, where the trading hours of different stock exchanges vary due to time zones, with partial overlap of different markets and very peculiar trading behavior at the beginning and the end of each trading day. Therefore foreign exchange rates are particularly suited for the study of long time series.

We consider the time series of hourly exchange rates between Euro and US dollar in the time period from January 2001 to May 2013. The empirical data were obtained from \texttt{www.fxhistoricaldata.com}\,. We denote the time series of exchange rates by $S(t)$. From these we calculate the time series of returns, i.e., the relative changes in the exchange rates on time intervals $\Delta t$,
\begin{equation}
r(t)=\frac{S(t+\Delta t)-S(t)}{S(t)} \ .
\end{equation}
Since we work with hourly data, the smallest possible value for $\Delta t$ is one hour. However, as we will see later on, a return interval of one trading day, $\Delta t=1 \mathrm{d}$, is preferable for the variance estimation.
Note that foreign exchange rates are typically modeled by a multiplicative random process, such as a geometric Brownian motion, see, \textit{e.g.}, \cite{biger1983valuation}. Therefore we consider the relative changes of the exchange rates instead of the exchange rates themselves. While the latter resemble -- at least locally - a lognormal distribution, the returns are approximately Gaussian, conditioned on the local variance, that is.

\section{Non-stationary variances in univariate time series}   \label{sec4}

\begin{figure}
\begin{center}
 \incfig{0.46\textwidth}{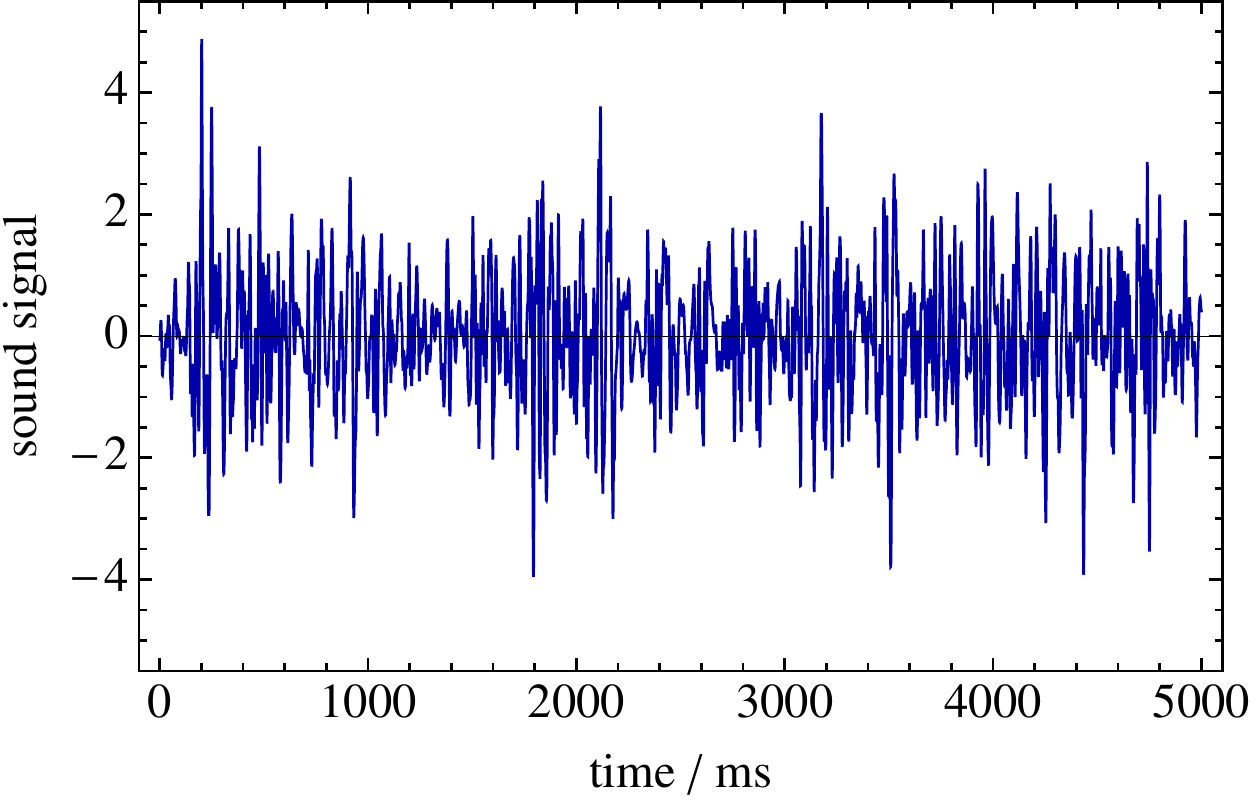} \\[2ex] 
 \incfig{0.46\textwidth}{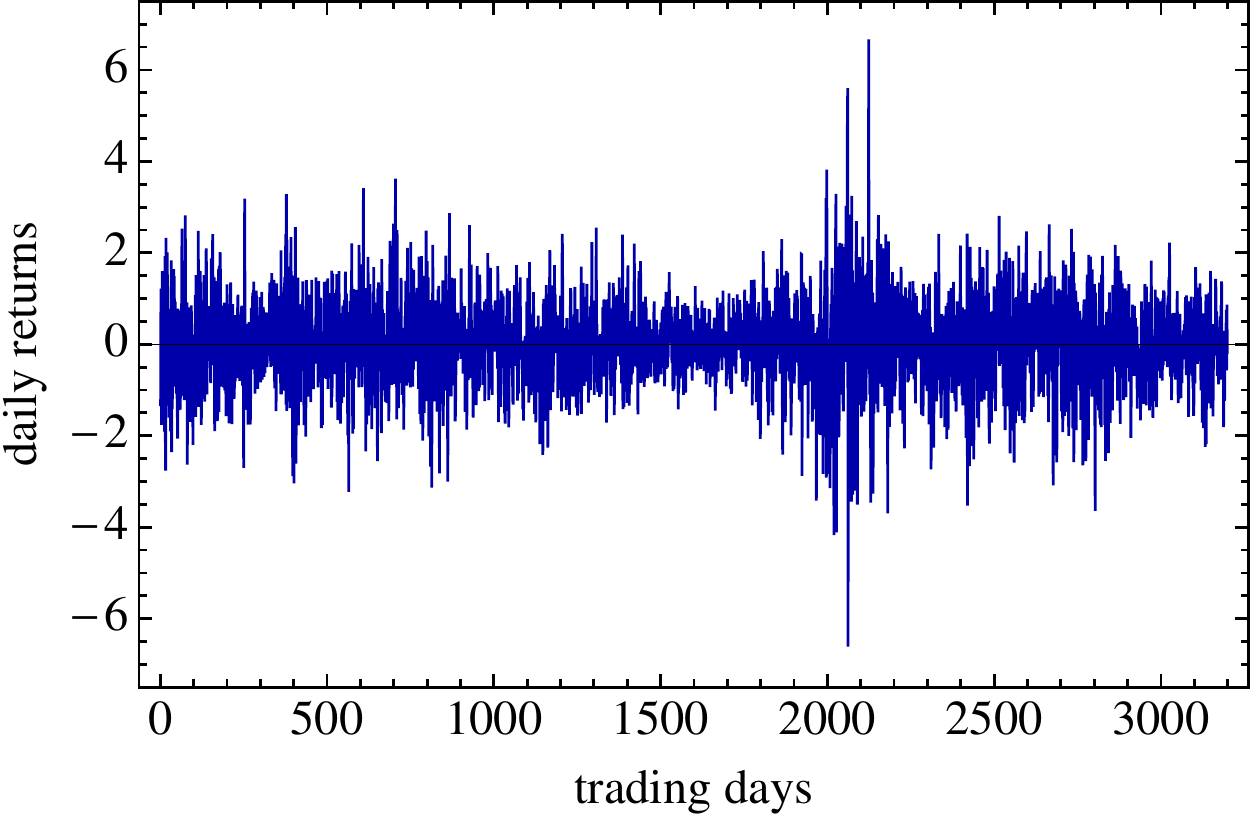}
\caption{
(Top) Sound signal of the ventilator measurement. (Bottom) Time series of daily returns. Both signals have been normalized to mean zero and standard deviation one. In both cases we observe extended periods of low and high fluctuation strength.}
\label{fig:signal}
\end{center}
\end{figure}

We consider the problem of univariate time series with time-dependent variance. More specifically, we consider time series where the variance is changing, but exhibits a slowly decaying autocorrelation function.
This point is crucial, because otherwise it is not possible to make meaningful estimates of the local variances.
Time series with this feature show extended periods of large fluctuations interupted by periods with moderate or small fluctuations. This is illustrated in Fig.~\ref{fig:signal} for the two data sets we are studying in this paper.
In the top plot of Fig.~\ref{fig:signal}, we show the sound signal for the ventilator measurement. In the bottom plot, the time series of daily returns for the foreign exchange data is plotted.
In both cases we observe the same qualitative behavior, which is well-known in the finance literature as volatility clustering.

The compounding ansatz for univariate time series assumes a normal distribution  on short time horizons, where the local variance is nearly stationary.
However, we wish to determine the distribution of the local variances empirically, since it is a critical part in the compounding ansatz.
If the variances were fluctuating without a noticable time-lagged correlation, this would not be feasible.
Still, we need to establish the right time horizon on which to estimate the local variances.
Ref.~\cite{SchaeferGuhr2010} introduced a method to locally normalize time series with autocorrelated variances.
To this end, a local average was subtracted from the data and the result was divided by a local standard deviation.
In this spirit, we determine the time horizon on which this local normalization yields normal distributed values and analyse the corresponding local variances.

\begin{figure}
\begin{center}
 \incfig{0.46\textwidth}{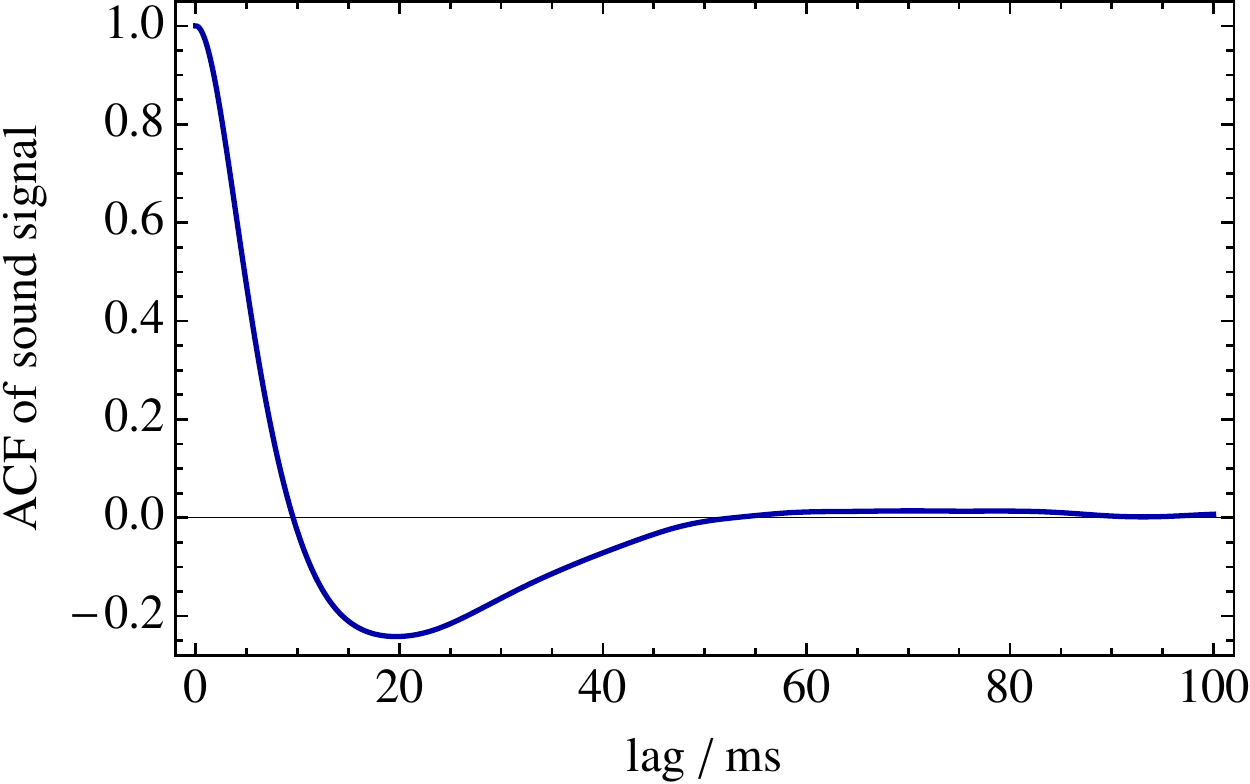} \\[2ex] 
 \incfig{0.46\textwidth}{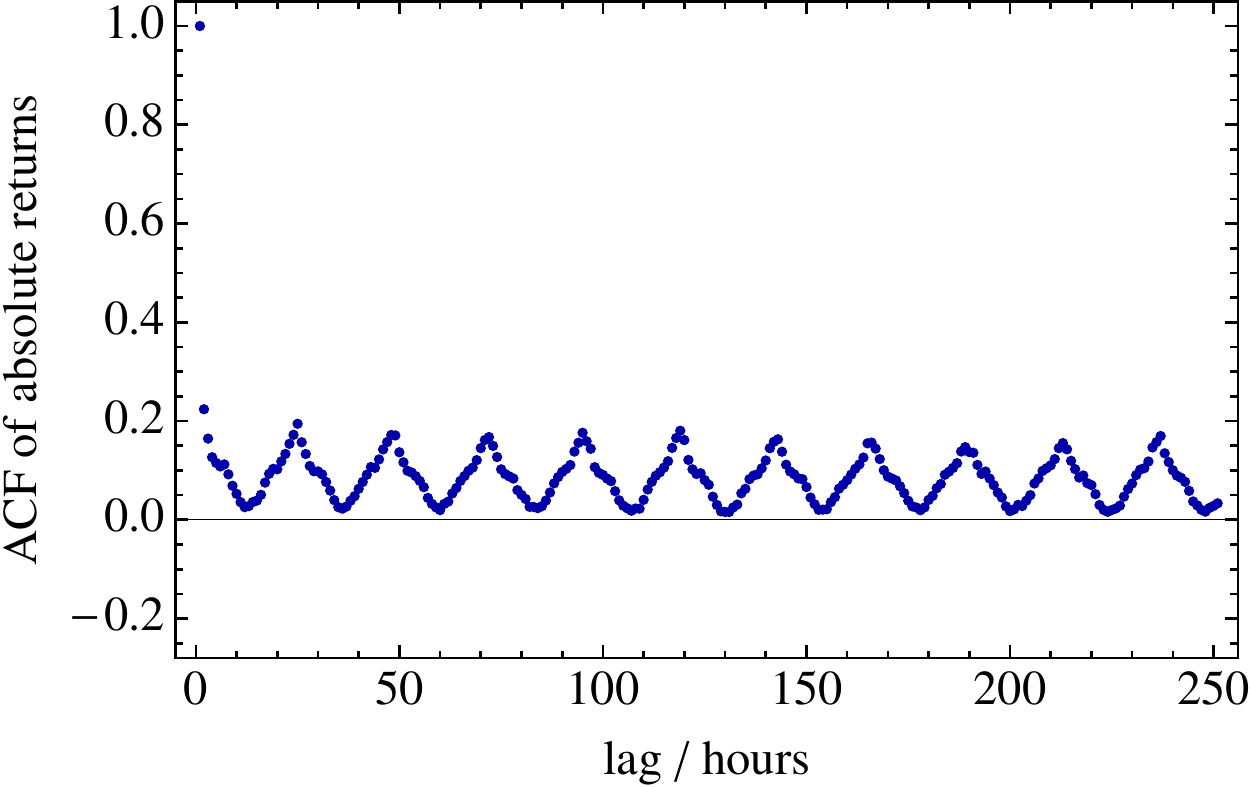}
\caption{
(Top) Autocorrelation function of the measured sound signal.
(Bottom) Autocorrelation function of the absolute values of hourly returns.}
\label{fig:acf}
\end{center}
\end{figure}

Another aspect we need to take into account is a possible bias in the variance estimation which occurs for correlated events.
In Fig.~\ref{fig:acf}  we show the autocorrelation function (ACF) of the measured sound signal, as well as the autocorrelation function of the absolute value of hourly returns.
Both plots hint at possible problems for the variance estimation.
Due to the high sampling frequency, the sound signal is highly correlated. 
In other words, the sampling time scale is much shorter than the time scale on which the turbulent air flow changes.
After 2500 data points, or about 52 ms, the autocorrelation function has decayed to zero. Consequently, we consider only every 2500th data point for our local variance estimation. 
To improve statistics, we repeat the variance estimation starting with an offset of 1 to 2499. The results are presented in the following section.

In the case of the foreign exchange data we are confronted with a different problem. The consecutive hourly returns are not correlated. While local trends may always exist, it is unpredictable when a positive trend switches to a negative one, and {\em vice versa}, see Ref.~\cite{preis2011switching}. 
However,  the autocorrelation of the absolute values shows a rich structure which is due to characteristic intraday variability. This would lead to a biased variance estimation and, consequently, to a distortion of the variance distribution. Therefore we consider returns between consecutive trading days at the same hour of the day. Put differently, we consider $\Delta t=1\,\mathrm{d}$ for the returns and get 24 different time series, one for each hour of the day as starting point.

\section{Empirical results}   \label{sec5}

\begin{figure}[t]
\begin{center}
 \incfig{0.46\textwidth}{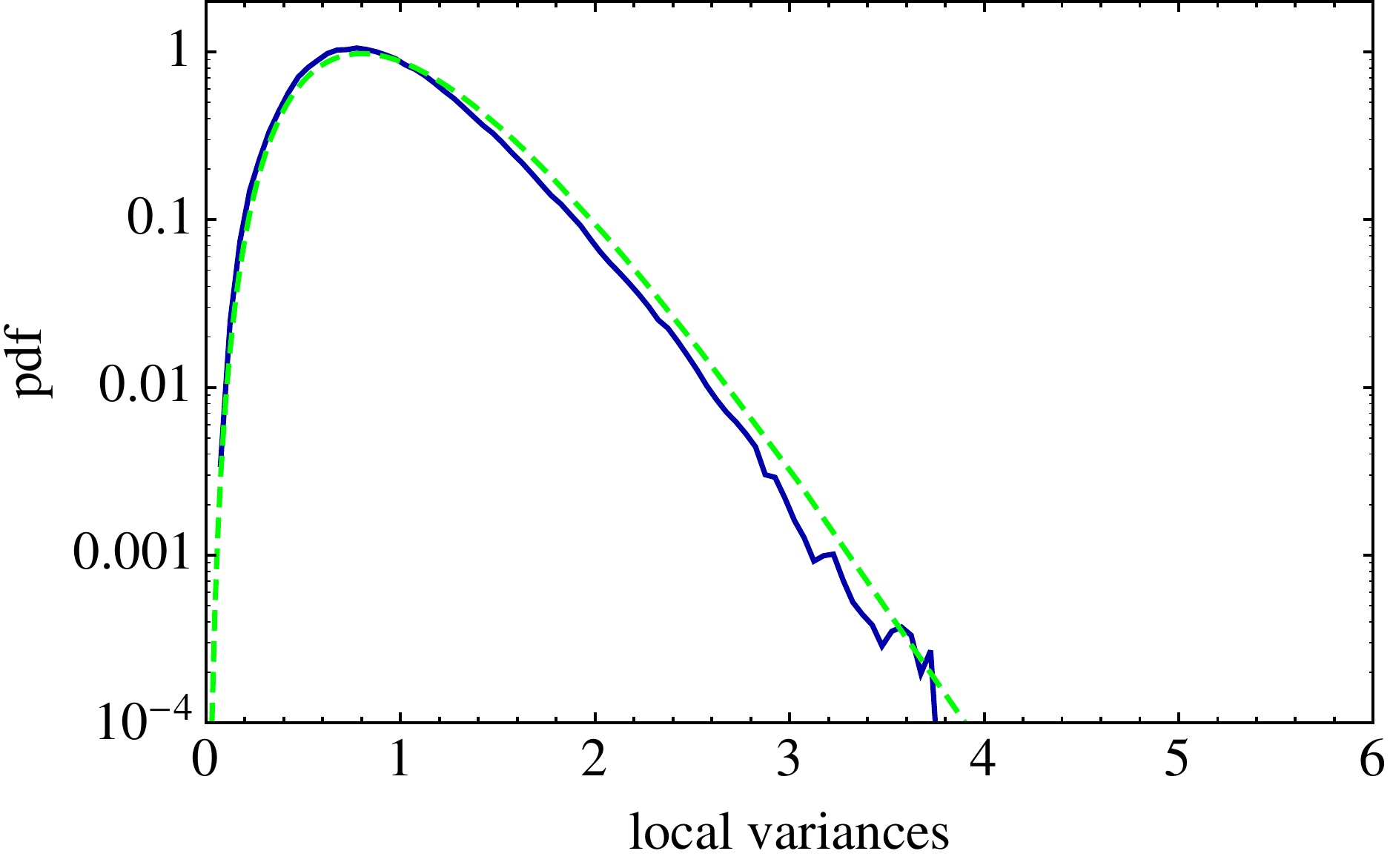} \\[2ex]
 \incfig{0.46\textwidth}{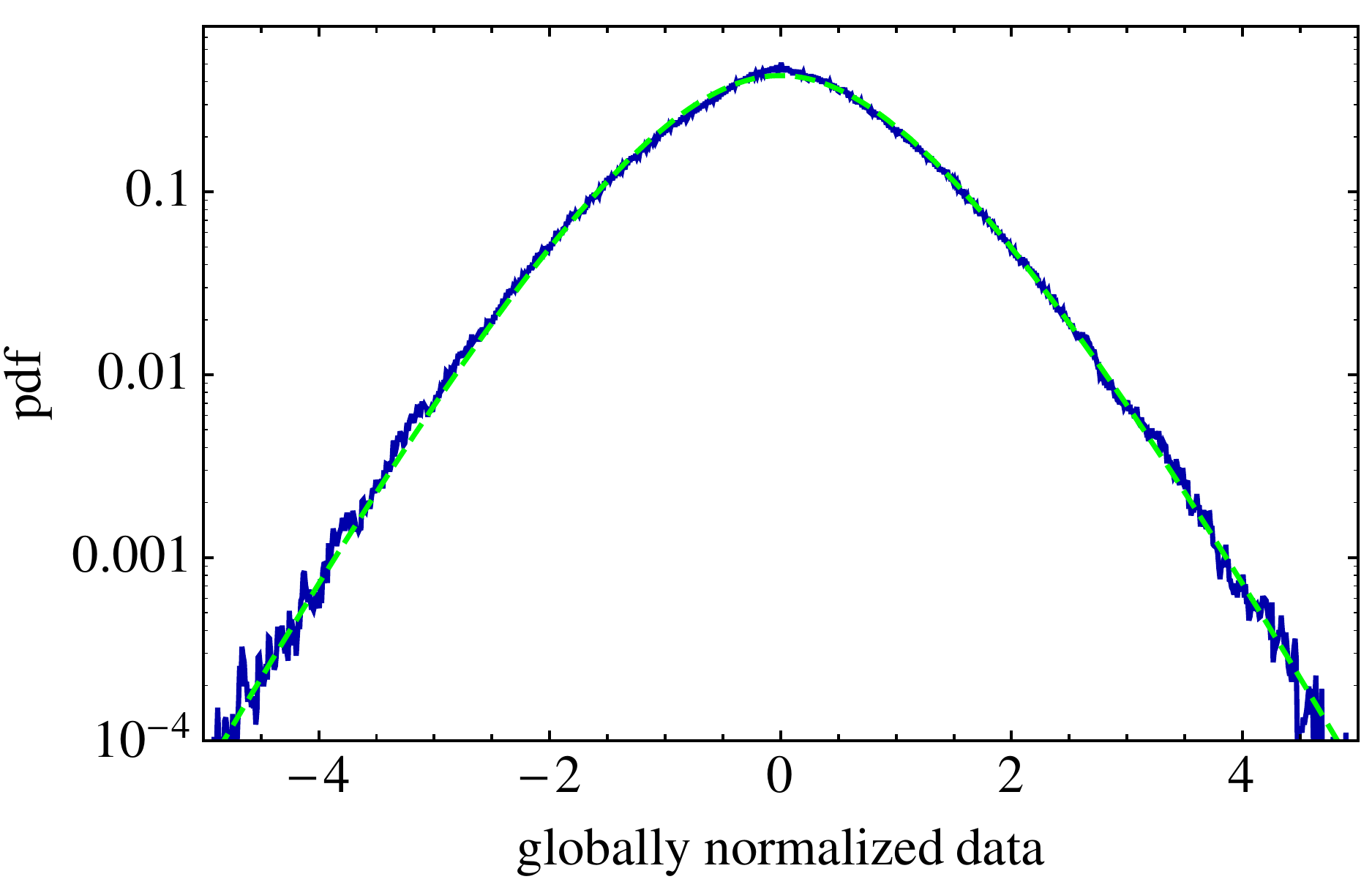}
\caption{
Results for the turbulent air flow. (Top) Distribution of variances, compared to a $\chi^2$-distribution with $N$ degrees of freedom.
(Bottom) Distribution of the sound amplitudes, compared to the K-distribution with parameter $N=10$.
}
\label{fig:venti}
\end{center}
\end{figure}

We first discuss the results for the turbulent air flow. 
As described in the previous section, we sliced the single measurement time series into 2500 time series with lower sampling rate, taking only every 2500th measurement point. This is necessary to avoid a bias in the estimation of the local variances. Before proceeding, each of these time series is globally normalized to mean zero and standard deviation one. 
Figure~\ref{fig:venti} shows the empirical results for the distribution of local variances and the compounded distribution. 
The local variances are rather well described by a $\chi^2$ distribution with $N$ degrees of freedom. We find $N=10$ to provide the best fit to the data. The distribution of the empirical sound amplitudes is well described by a 
K-distribution with the same $N$ which fits the variance distribution. Hence, we arrive at a consistent picture, which supports our compounding ansatz for this measurement.

\begin{figure}[t]
\begin{center}
 \incfig{0.46\textwidth}{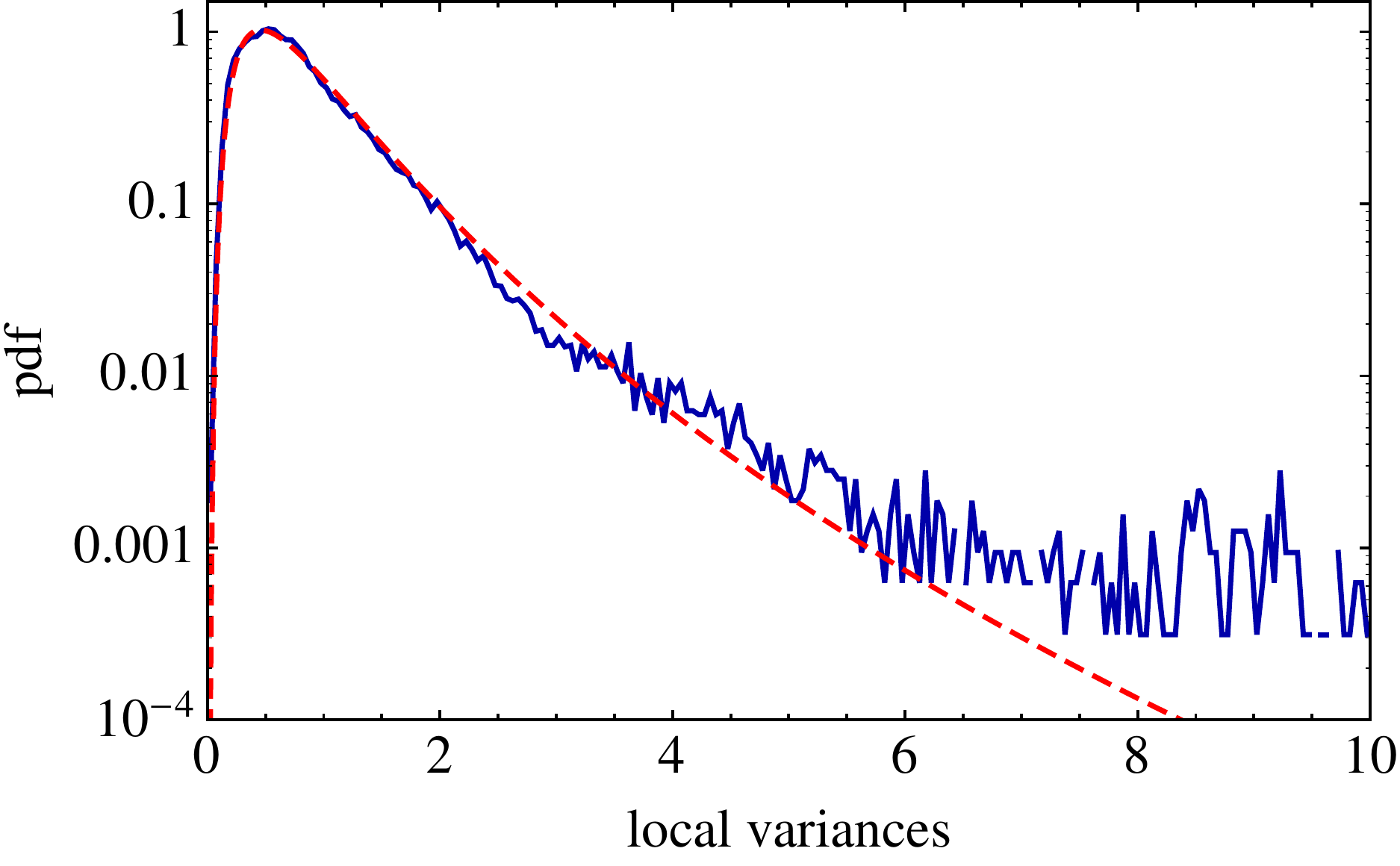} \\[2ex]
 \incfig{0.46\textwidth}{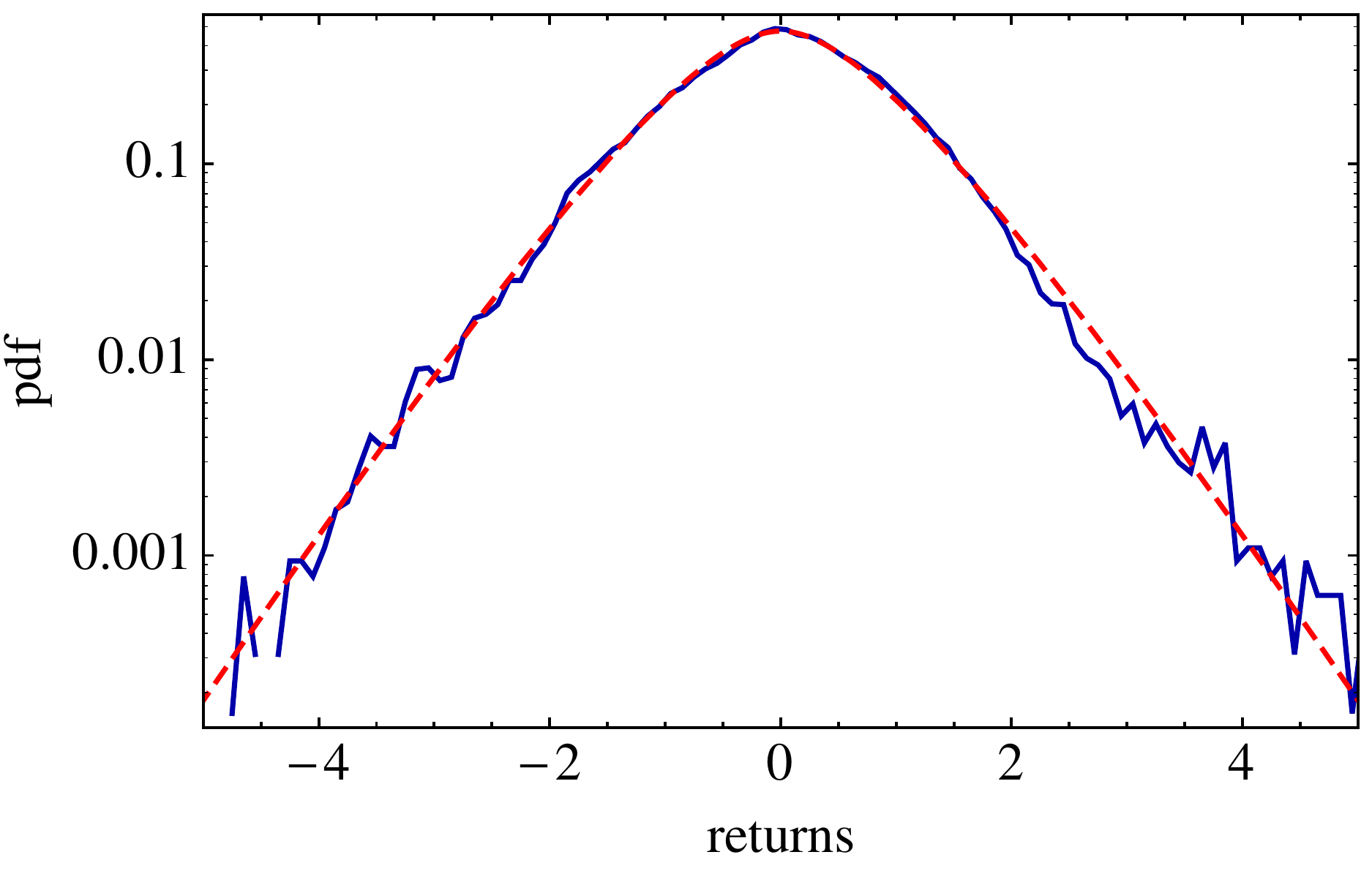}
\caption{
Results for daily returns of foreign exchange rates EUR-USD. (Top) Distribution of variances, compared to a lognormal distribution.
(Bottom) Distribution of the daily returns, compared to a lognormal compounded distribution (red). 
}
\label{fig:forex}
\end{center}
\end{figure}

The results for the daily returns of EUR-USD foreign exchange rates are shown in Fig.~\ref{fig:forex}.
As outlined in section \ref{sec4}, we calculated the daily returns as the relative changes of the exchange rate between consecutive trading days with respect to the same hour of each day.  This procedure yields 24 time series of daily returns.
We normalize each time series to mean zero and standard deviation one. This allows us to produce a single aggregated statistics.
In the top plot of Fig.~\ref{fig:forex} we show the histogram of local variances, i.e. the variances estimated on 13-day intervals.
In accordance with the finance literature, the empirical variances follow a lognormal distribution over almost three orders of magnitude, with only some deviations in the tail.
The histogram of the daily returns is shown in the bottom plot of Fig.~\ref{fig:forex}. The empirical result agrees rather well with the normal-lognormal compounded distribution.
It is important to note, however, that we only achieve this consistent picture of variance and compounded return distribution because we have taken into account all the pitfalls of variance estimation, which we described in section \ref{sec4}.

\section{Conclusions}   \label{con}

We applied the compounding approach to two different systems, a ventilator setup generating turbulent air flow and foreign exchange rates.
Both systems are characterized by univariate time series with non-stationary variances.
Our main objective was to empirically determine the distribution of variances and thus arrive at a consistent picture.
The estimation of variances from a single, non-stationary time series presents several pitfalls, which have to be taken into account carefully. First of all, we have to avoid serial correlations in the signal itself. These might otherwise lead to an estimation bias. 
For the sound measurement, we had to reduce the sampling rate of the data to achieve this. 
The foreign exchange data presented another obstacle for variance estimation: We observed a characteristic intraday variability which had to be taken into account.
Last but not least, it is a prerequisite that the non-stationary variances are not purely stochastic, but exhibit a slowly decaying autocorrelation. Otherwise we would not be able to determine a reasonable variance distribution for the compounding ansatz.
When we take all these aspects into account, we arrive at the correct variance distribution. In good approximation we found a $\chi^2$ distribution in the case of ventilator turbulence, which leads to a K-distribution for the compounded statistics. For the foreign exchange returns we observe lognormal distributed variances; and the normal-lognormal compounded distribution fit the return histogram well.
A central assumption in the compounding ansatz is the stationarity of the variance distribution. This assumption might not always be satisfied and lead to deviations from the compounded distribution.

\section*{Acknowledgments}

The sound measurements have been supported by the Deutsche Forschungsgemeinschaft via the Forschergruppe 760  ``Scattering Systems with Complex Dynamics''.


\end{document}